\documentstyle[11pt]{article}
\begin{document}

\title{Mean Field in Long-Range Ferromagnets and Periodic Boundary Conditions}
\author{ Sergio Curilef \\
\em Departamento de F{\'\i}sica, Universidad Cat\'olica del Norte, \\
\em Casilla 1280, Antofagasta, Chile\\
E-mail: scurilef@ucn.cl}
\maketitle

\begin{abstract}
Periodic boundary conditions are applied to a ferromagnetic spin lattice.
A symmetrical lattice and its contributions all over space are being used.
Results, for  
the Ising model with ferromagnetic interaction that decays as a $1/r^{D+\nu}$ law,
are discussed in the mean field approximation.   \\

\end{abstract}

Thermodynamics describes the behavior of systems with many degrees of 
freedom after 
they have reached a state of thermal equilibrium. However, their thermodynamic state
can be specified in terms of a few parameters, they are called extensive or intensive state variables. 
At equilibrium, macroscopic observables are linear functions of 
$N$ (number of particles). If the function $f(N)$ is an observable, the extensive property imposes 
$f(N) = Nf_N $, where $f_N = f(N)/N$ and $f(\lambda N) = \lambda f(N)$ for very large $N$\cite{fisher}.

Small size of system are considered for studying thermodynamic behavior with
computer simulations, typically $N \sim 500 $ particles. Available memory 
and speed of program execution (in the host computer) can limit the size of 
the simulated systems.
However, in thermodynamics is necessary to obtain the trend of the thermodynamic
properties for very large systems, $N \rightarrow \infty$. 
Then, how can we expect to simulate with finite systems some properties 
related to infinite systems? The answer for this question depends on the property under investigation 
and the range of microscopic interparticle interactions.

Periodic summations of forces do not depend on the nature of interactions, 
even for long range interactions, forces converge very fast\cite{SCuIJM11}.
If the range of interaction is smaller than the size of the system, computer 
simulations give the typical trends of the thermodynamic behavior of systems.
When interactions are long-ranged, surface effects appear and become very important 
for all system sizes. A first approach requires to increase to very large size of the box. 

A long-range interaction is often defined as one in which the space interactions falls of no 
faster than $1/r^D$, 
where $D$ is the space dimension of the system.

To eliminate surface effects from the computation let us use a particular procedure. 
Surface effects can be ignored for all system sizes if we use periodic boundary conditions. 
In periodic boundary conditions, a central cell is replicated throughout space to form an 
infinite lattice. In the course of the simulation, when a molecule moves in the central cell, 
its periodic image in every one of the other cells moves with exactly the same orientation 
in exactly the same way\cite{SCuIJM11,SCuPLA299,JLePhA176,NGrIJM6}. Thus, as a molecule leaves the central cell, one of its images 
will enter through the opposite face. There are no walls at the boundary of the central cell, 
and the system has no surface. 
The central cell simply forms a convenient coordinate system for measuring locations of the N
particles. 

Therefore, the size of the systems should depend on two parameters, 
${\cal N}$ and  $M$, as follows
\begin{equation}
N = (2M+1)^D {\cal N}
\label{calN}
\end{equation}
where ${\cal N}$ is the number of particles in every cell and $(2M+1)^D$ represents 
the number of replications on all space. So, thermodynamic limit ($N \rightarrow \infty$)
is reached when ${\cal N} \rightarrow \infty$, but this solution is impracticable even with
modern computers, the time taken to evaluate the energy increases proportional 
to ${\cal N}^2$. An alternative way to approach the thermodynamic limit is to take a finite 
value of ${\cal N}$ and $M \rightarrow \infty$.

In recent years, due to the difficult to obtain exact solutions in systems with 
long-range interactions, much effort has been devoted to handling this kind of interactions
in computational systems by molecular dynamics and Monte Carlo simulations. 
Another reason for developing simulation techniques is to test some approximated results\cite{HasPRL91}.
In the present work, a long-range Ising lattice is studied in the mean field 
approximation. Furthermore, we look for a cutoff for involved summations in thermodynamics 
by a nonextensive scaling. 

We consider a computational cell in $D$
dimensions (normally $D$=1,2,3) with size $N$ and 
periodic boundary conditions given by repetition of a central cell to
infinity. 
The central cell contains ${\cal N}$ spin$\frac{1}{2}$ Ising ferromagnets with two-particle 
potential\cite{SCuPLA299,SCaPRB54,SCaPRB61}
and magnetic moment $\mu$  that means, a
system described by the Hamiltonian
\begin{equation}
H = - \sum_{(i,j)} J(\vec{r}_{ij})S_iS_j - \mu h \sum_i S_i 
\end{equation}
where $h$ is an external magnetic field, $S_i= \pm 1 \;\;\;\forall i$ and $r_{ij}$ is the distance between two sites and
\begin{equation}
J(\vec{r}) = J \sum^{\vec{M}}_{\vec{K}=-\vec{M}} \frac{1}{|\vec{r}+\vec{K}|^{D+\nu}}
\end{equation}
where $J>0$ measures the strength of the coupling. In precedent works, topics of systems with long-range 
interactions have been discussed, by several authors\cite{HasPRL91,BVoPRE63}.
A common behavior has been
conjectured~\cite{TsaFRA95} for generic systems with arbitrary (long or short) range 
interactions. In this way, it is defined for continuos variations
$N^* \equiv (N^{-\nu/D} -1)/(-\nu/D)$~\cite{TsaFRA95} and for a spin lattice
$\bar{N} \equiv 2^{D+\nu} N^*$~\cite{SCaPRB54}. Now, we can expect that quantities 
like internal energy, free energy, etc., per particle scale with  $N^*$ and $\bar N$, respectively . 
In the present case, variables include scaling with $\bar N$; more explicitly,
the so called intensive variables $(T,P,$ etc.) scale with $\bar N$
Now, we calculate average of hamiltonian summing over the configuration space, 
the dependence on $N^2$ appears for an approximated evaluation of energy.
Close to mean field approximation,  the hamiltonian of a $D$-dimensional spin lattice 
with $N$ lattice sites can be written

\begin{equation}
H = - \sum_{i=1}^N E(\bar N,h)S_i  
\end{equation}
where $E(\bar N,h) = - \frac{1}{2} \bar N \langle S \rangle - \mu h$, the factor $\frac{1}{2}$ 
ensures that we do not count the same pair of spin twice. The quantity $\langle S \rangle$ is 
the average spin per site. 
This assumption helps us to recover all the results for  system with long range microscopic interactions 
in a similar manner to the traditional mean field approximation.

So, the average spin per site of the lattice is given by  
\begin{equation}
\langle S \rangle = \tanh \left[ \beta \left(\frac{1}{2} \bar N J \langle S \rangle +\mu h\right) \right]
\end{equation}
If $h=0$, the magnetization $N\mu\langle S \rangle$ will be zero for high temperature paramagnetic phase 
and it will be nonzero at lower temperatures where the spins have spontaneously aligned.
This transition (critical) point is obtained when $ \bar N J / 2 k_B T = 1$. So, the critical 
temperature closes to the mean field value
\begin{equation} T_c = \bar N J / 2 k_B \end{equation}
where $T_c/{\bar N}$ takes a constant value, that does not depend on the range of the microscopic
interaction in the present approximation. 

In a simple numerical test for the critical point of a long-range Ising ferromagnet system , in one dimension, 
is obtained a departure from mean field approximation. We note that if 
$\nu \rightarrow -1$, the mean field model closes to the present system; but if $\nu > -1$, simulated 
critical temperature $T_c/{\bar N}$  is less than the mean field value. 
We have performed numerical calculations to test different ranges of parameters for system with long-range
interactions and a more detailed study of critical phenomena for systems with this kind of microscopic interactions
is in advance\cite{preprint}. 

The internal energy is given by

\begin{equation} 
U(N) = -\frac{1}{2} N\bar N J\langle S \rangle^2 .
\label{UN}  
\end{equation}
Now, we can replace Eq.(\ref{calN}) in Eq.(\ref{UN})
\begin{equation} 
U((2M+1)^D {\cal N}) = \frac{D}{|\nu|}(2M+1)^{D+|\nu|} U({\cal N}),  
\end{equation}
where $\nu < 0$. So, $U(N)$ violates the extensive property and it is a homogeneous function of 
degree $(1+ |\nu|/D)$. 

The heat capacity becomes
\begin{equation}
C = \frac{2Nk_B\langle S \rangle^2 (T_c/T)^2}{\cosh^2(\langle S \rangle T_c/T)-T_c/T}
\end{equation}
and the magnetic susceptibility in the limit $h \rightarrow 0$ is
\begin{equation}
\chi = \frac{N}{{\bar N}} \frac{2k_B}{J}\frac{(T_c/T)}{\cosh^2(\langle S \rangle T_c/T)-T_c/T}
\end{equation}
Finally, it is important to understand the microscopic behavior of the
particles, because computation of the mesoscopic and macroscopic
thermodynamical properties
comes from the averages of quantities per particle when the number of
particles increases to infinity.  
Advances in calculation of these typical and physical quantities is very important in the study of 
phase transitions and critical phenomena for this kind of systems. 

This work has received partial support by FONDECYT, grant 1010776.

\end{document}